\documentclass[11pt]{article}
\usepackage{geometry}
\usepackage{amsfonts}
\usepackage{amsmath}
\usepackage{amssymb}
\usepackage{mathptmx}
\usepackage{color}
\usepackage{epsfig}
\usepackage{graphicx}
\usepackage{hyperref}
\usepackage{cancel}
\usepackage{textcomp}
\usepackage{bbold}
\usepackage{bm}
\usepackage{times}
\usepackage{slashed}
\usepackage{booktabs}
\usepackage{verbatim}
\usepackage{latexsym}
\usepackage{extarrows}
\usepackage{multirow}
\usepackage[dvipsnames]{xcolor}

\newcommand{\Tr}{\rm Tr}

\newcommand{\calL}{\mathcal{L}}
\newcommand{\calO}{\mathcal{O}}

\begin{document}
\begin{center}
{\Large\sf Perturbative power counting, lowest-index operators and their renormalization in standard model effective field theory}
\\[10pt]
\vspace{.5 cm}
{Yi Liao~$^{a,b,c}$\footnote{liaoy@nankai.edu.cn} and Xiao-Dong Ma~$^{a}$\footnote{maxid@mail.nankai.edu.cn}}

{$^a$~School of Physics, Nankai University, Tianjin 300071, China
\\
$^b$ CAS Key Laboratory of Theoretical Physics, Institute of Theoretical Physics,
Chinese Academy of Sciences, Beijing~100190, China
\\
$^c$ Synergetic Innovation Center for Quantum Effects and Applications,
Hunan Normal University, Changsha, Hunan~410081, China}

\vspace{2.0ex}

{\bf Abstract}

\end{center}

We study two aspects of higher dimensional operators in standard model effective field theory. We first introduce a perturbative power counting rule for the entries in the anomalous dimension matrix of operators with equal mass dimension. The power counting is determined by the number of loops and the difference of the indices of the two operators involved, which in turn is defined by assuming that all terms in the standard model Lagrangian have an equal perturbative power. Then we show that the operators with the lowest index are unique at each mass dimension $d$, i.e., $(H^\dagger H)^{d/2}$ for even $d\geq 4$, and $(L^T\epsilon H)C(L^T\epsilon H)^T(H^\dagger H)^{(d-5)/2}$ for odd $d\geq 5$. Here $H,~L$ are the Higgs and lepton doublet, and $\epsilon,~C$ the antisymmetric matrix of rank two and the charge conjugation matrix, respectively. The renormalization group running of these operators can be studied separately from other operators of equal mass dimension at the leading order in power counting. We compute their anomalous dimensions at one loop for general $d$ and find that they are enhanced quadratically in $d$ due to combinatorics. We also make connections with classification of operators in terms of their holomorphic and anti-holomorphic weights.

\newpage

We study in this short paper two general aspects in standard model effective field theory (SMEFT). One is a power counting rule in perturbation theory for anomalous dimension matrix of higher dimensional operators with equal mass (canonical) dimension that is induced by the standard model (SM) interactions. We show that the leading power of each entry in the anomalous dimension matrix is determined in terms of the loop order and the difference of indices for the two operators involved. The other is about the lowest-index operators. We find that they are unique at each dimension and can be renormalized independently of other operators of equal dimension at the leading order in SM interactions. We compute their one-loop anomalous dimensions, and find that they increase quadratically with their dimension due to combinatorics.

Regarding SM as an effective field theory below the electroweak scale, the low energy effects of high scale physics can be parameterized in terms of higher dimensional operators:
\begin{equation}
\mathcal{L}_\textrm{SMEFT}=\mathcal{L}_4+\mathcal{L}_5+\mathcal{L}_6+\mathcal{L}_7+\cdots.
\label{LSMEFT}
\end{equation}
Here the leading terms are the SM Lagrangian
\begin{eqnarray}
\label{sml}
\nonumber
\mathcal{L}_4&=&
-\frac{1}{4}\sum_X X_{\mu\nu}X^{\mu\nu}+(D_\mu H)^\dagger(D^\mu H)
-\lambda\left(H^\dagger H-\frac{1}{2}v^2\right)^2
\\
&&
+\sum_\Psi\bar{\Psi}i \slashed{D}\Psi
-\left[\bar{Q}Y_u u \tilde{H}+\bar{Q}Y_d d H+\bar{L}Y_e e H +\mbox{h.c.}\right],
\end{eqnarray}
where $X$ sums over the three gauge field strengths of couplings $g_{1,2,3}$, and $\Psi$ extends over the lepton and quark left-handed doublets $L,~Q$ and right-handed singlets $e,~u,~d$. The Higgs field $H$ develops the vacuum expectation value $v/\sqrt{2}$, and $\tilde H_i=\epsilon_{ij}H^*_j$. $D_\mu$ is the usual gauge covariant derivative, and $Y_{u,d,e}$ are Yukawa coupling matrices.

The higher dimensional operators, collected in $\calL_{5,6,7}$ and ellipses in Eq.~(\ref{LSMEFT}), are composed of the above SM fields, and respect the SM gauge symmetries but not necessarily accident symmetries like lepton or baryon number conservation. They are generated from high scale physics by integrating out heavy degrees of freedom, with their Wilson coefficients naturally suppressed by powers of certain high scale. It is thus consistent to leave aside those Wilson coefficients when we do power counting for their renormalization running effects due to SM interactions. The higher dimensional operators start at dimension-five (dim-5), which turns out to be unique~\cite{Weinberg:1979sa}. The complete and independent list of dim-6 and dim-7 operators has been constructed in Refs.~\cite{Buchmuller:1985jz, Grzadkowski:2010es} and \cite{Lehman:2014jma,Liao:2016hru} respectively. The number of operators increases horribly fast with their dimension; for discussions on dim-8 operators and beyond, see recent papers \cite{Lehman:2015via,Henning:2015daa,Lehman:2015coa,Henning:2015alf}.
If SM is augmented by sterile neutrinos below the electroweak scale, there will be additional operators at each dimension, see Refs.~\cite{Aparici:2009fh,delAguila:2008ir, Bhattacharya:2015vja,Liao:2016qyd} for discussions on operators up to dim-7 that involve sterile neutrinos.

Now we consider power counting in the anomalous dimension matrix $\gamma$ of higher dimensional operators due to SM interactions. We restrict ourselves in this work to the mixing of operators with equal mass dimension, because this is the leading renormalization effect due to SM interactions that is not suppressed by a high scale. Since the power counting is additive, it is natural to assign an index of power counting $\chi[\calO]$ to the operator $\calO$ which in turn is a sum of the indices for the elements involved in $\calO$. For the purpose of power counting, we denote $g$ as a generic coupling in SM. Suppose an effective interaction $C_i\calO_i$ in $\calL_\textrm{SMEFT}$ is dressed by SM interactions at $n$-loops to induce an effective interaction, $\Delta_{ji}\calO_j$ (no sum over $j$), involving the operator $\calO_j$ of equal dimension. The SM $n$-loop factor of $g^{2n}$ is shared by the difference of the indices of the operators $\chi[\calO_j]-\chi[\calO_i]$ and the induced ultraviolate divergent coefficient $\Delta_{ji}$. As $\Delta_{ji}\calO_j$ contributes a counterterm to the effective interaction $C_j\calO_j$ from which $\gamma_{ji}$ is determined for the running of $C_j$, we obtain the power counting for the entry $\gamma_{ji}$ in the anomalous dimension matrix
\begin{equation}
\chi[\gamma_{ji}]=2n+\chi[\calO_i]-\chi[\calO_j].\
\label{chi_gamma}
\end{equation}
The issue now becomes defining an index for operators up to a constant, $\chi[\calO]$, which could be understood as an intrinsic power counting of SM couplings for the operator $\calO$.

Since we are concerned with overall power counting in SM interactions, it is plausible to treat all terms in $\calL_4$ on the same footing by assuming an equal index of perturbative power counting when the kinetic terms have been canonically normalized.
A similar argument was assumed previously in chiral perturbation theory involving chiral fermions coupled to electromagnetism~\cite{Urech:1994hd,Knecht:1999ag,Nyffeler:1999ap,chiralNDA}. Denoting generically
\begin{equation}
\chi[H]=x,~\chi[\lambda]=2y,
\end{equation}
so that $\chi[\calL_4]=4x+2y$, it is straightforward to determine the indices of other components in $\calL_4$:
\begin{equation}
\chi[\Psi]=\frac{3}{2}x+\frac{1}{2}y,~\chi[X_{\mu\nu}]=2x+y,
~\chi[D_\mu]=x+y,~\chi[g_{1,2,3}]=\chi[Y]=y.
\end{equation}
It is evident that the $x$ term actually counts canonical dimension and the $y$ counts the power of $g$. Since we are concerned with renormalization mixing of operators with equal dimension, the power counting for their anomalous dimension matrix depends only on the $y$ term according to Eq.~(\ref{chi_gamma}). Although our $\chi[\gamma_{ij}]$ does not depend on $x$, we find it most convenient to work with $x=0$ and $y=1$, so that the nonvanishing indices for power counting are
\begin{equation}\label{count_basic}
\chi[\Psi]=\frac{1}{2},~\chi[X_{\mu\nu}]=1,~\chi[D_\mu]=1,~\chi[g_{1,2,3}]=\chi[Y]=1,~\chi[\lambda]=2.
\end{equation}
The lowest index that an operator could have is zero in this convention. Using a different $x$ amounts to shifting the indices of all fields and derivatives by a multiplier of their mass dimensions without changing $\chi[\gamma_{ij}]$, and choosing $y=1$ simply fits the usual convention that all gauge and Yukawa couplings count as $g^1$ while the scalar self-coupling $\lambda$ counts as a quartic gauge coupling $g^2$.

\begin{table}
\center
\begin{tabular}{c c c c c c c c  c c}
\hline
& index   & $3$ & $2$ & $2$ & $2$ & $2$ & $2$ & $1$ & $0$ \\
\hline
index &  $\gamma_{ij}$      & $X^3$    & $X^2H^2$       & $\bar{\Psi}\Psi HX$        & $H^4D^2$      & $\bar{\Psi}\Psi H^2D$     & $\bar{\Psi}^2\Psi^2$  & $\bar{\Psi}\Psi H^3$     & $H^6$
\\
\hline
$3$ & $X^3$  & $g^2$ & $g^1$ & $g^1$ & $g^1$ & $g^1$ & $g^1$ & 0 & 0
\\
\hline
$2$ & $X^2H^2$ & $g^3$& $g^2$ & $g^2$  & $g^2$ &$g^2$ & $g^2$ & $g^1$ & 0
\\
\hline
$2$ & $\bar{\Psi}\Psi HX$ & $g^3$& $g^2$ & $g^2$  & $g^2$ &$g^2$ & $g^2$ & $g^1$ & 0
\\
\hline
$2$ & $H^4D^2$  & $g^3$& $g^2$ & $g^2$  & $g^2$ &$g^2$ & $g^2$ & $g^1$ & 0
\\
\hline
$2$ & $\bar{\Psi}\Psi H^2D$ & $g^3$& $g^2$ & $g^2$  & $g^2$ &$g^2$ & $g^2$ & $g^1$ & 0
\\
\hline
$2$& $\bar{\Psi}^2\Psi^2$  & $g^3$& $g^2$ & $g^2$  & $g^2$ &$g^2$ & $g^2$ & $g^1$ & 0
\\
\hline
$1$& $\bar{\Psi}\Psi H^3$  &  $g^4$ & $g^3$ & $g^3$  & $g^3$ &$g^3$ & $g^3$ &  $g^2$  & $g^1$
\\
\hline
$0$&  $H^6$ & $g^5$  & $g^4$  & $g^4$ & $g^4$ &  $g^4$ &   $g^4$  & $g^3$  & $g^2$
\\
\hline
\end{tabular}
\caption{Indices of power counting for dim-6 operators and power counting of their anomalous dimension matrix at one loop.}
\label{tab_dim6}
\end{table}

\begin{table}
\center
\begin{tabular}{c c c c c c c c}
\hline
& index   & $3$ & $3$ & $2$ & $2$ & $2$ & $1$\\
\hline
index &  $\gamma_{ij}$      & $\Psi^2H^2D^2$    & $\bar{\Psi}\Psi^3D$       & $\Psi^2H^2X$        & $\Psi^2H^3D$      & $\bar{\Psi}\Psi^3H$     & $\Psi^2H^4$
\\
\hline
$3$ &$\Psi^2H^2D^2$   & $g^2$& $g^2$ & $g^1$ & $g^1$ & $g^1$ &0
\\
\hline
$3$ & $\bar{\Psi}\Psi^3D$   & $g^2$& $g^2$ & $g^1$ & $g^1$ & $g^1$ & 0
\\
\hline
$2$ &$\Psi^2H^2X$   & $g^3$  & $g^3$ & $g^2$  & $g^2$  & $g^2$ & $g^1$
\\
\hline
$2$ &$\Psi^2H^3D$   & $g^3$ & $g^3$ & $g^2$   & $g^2$  & $g^2$ & $g^1$
\\
\hline
$2$ &$\bar{\Psi}\Psi^3H$ & $g^3$ & $g^3$ & $g^2$ & $g^2$ & $g^2$ & $g^1$
\\
\hline
$1$&$\Psi^2H^4$ & $g^4$ & $g^4$  & $g^3$& $g^3$ & $g^3$ & $g^2$
\\
\hline
\end{tabular}
\caption{Similar to Table~\ref{tab_dim6} but for dim-7 operators. }
\label{tab_dim7}
\end{table}

We can now associate an index of power counting $\chi[\calO]$ to a higher dimensional operator $\calO$ by simply adding up the indices of its components according to Eq.~(\ref{count_basic}). The entry $\gamma_{ji}$ in the anomalous dimension matrix for the set of operators $\calO_k$ due to SM interactions at $n$-loops has the index of power counting shown in Eq.~(\ref{chi_gamma}) in terms of a generic coupling $g$, which denotes $g_{1,2,3}$, $Y_{e,u,d}$, and $\sqrt{\lambda}$. Our results for dim-6 and dim-7 operators are shown in Table~\ref{tab_dim6} and Table~\ref{tab_dim7} respectively. The one-loop $\gamma$ matrix for dim-6 operators has been computed in a series of papers~\cite{Grojean:2013kd,Elias-Miro:2013gya,Elias-Miro:2013mua,Jenkins:2013zja,
Jenkins:2013wua,Alonso:2013hga,Alonso:2014zka}, and is consistent with power counting in Table~\ref{tab_dim6}. The $\gamma$ submatrix for baryon number violating dim-7 operators is available recently~\cite{Liao:2016hru}, and also matches power counting in Table~\ref{tab_dim7}. Note that some entries in the tables may actually vanish due to structures of one-loop Feynman diagrams or nonrenormalization theorem \cite{Alonso:2014rga,Elias-Miro:2014eia,Cheung:2015aba}. Since at least one vertex of SM interactions is involved in one-loop diagrams, $\gamma$ counts as $g^1$ or higher. This explains the presence of zero in the last two columns of the tables. The power counting in the explicit result of one-loop $\gamma$ matrix for dim-6 operators has also been explained in Ref.~\cite{Jenkins:2013sda} using the arguments of naive dimensional analysis developed for strong dynamics~\cite{Manohar:1983md} that rescale operators forth and back by factors of couplings and powers of $4\pi$. Our analysis above is more straightforward and assumes only the uniform application of SM perturbation theory.

With the above definition of the index of power counting for an operator, we make an interesting observation that the operator with the lowest index is unique at each mass dimension. To show this, we notice that out of the building blocks ($H,~\Psi,~D_\mu,~X_{\mu\nu}$) for higher dimensional operators only $H$ has a vanishing index. This means that it should appear as many times as possible in the lowest-index operators for a given mass dimension $d$. For $d$ even, this is easy to figure out, i.e.,
\begin{eqnarray}
\calO^d_H&=&(H^\dagger H)^{d/2}.
\label{O_H}
\end{eqnarray}
These operators represent a correction to the SM scalar potential from high scale physics, and could impact the vacuum properties. For $d$ odd, additional building blocks must be introduced. In the absence of fermions, $X_{\mu\nu}$ and $D_\mu$ have to appear at least twice due to Lorentz invariance, which costs no less than two units of index. And in addition, this cannot yield an operator of odd dimension. The cheapest possible way would be to introduce two fermion fields in a scalar bilinear form on top of the Higgs fields, resulting in an operator of index unity. It turns out that gauge symmetries require the fermions to be leptons. Sorting out the quantum numbers of lepton fields \footnote{The bilinear form $(\bar{L}e)$ must couple to an odd total number of $H^\dagger$ and $H$ thus resulting in an even dim-$d$ operator. The bilinear $(ee)$ requires four more powers of $H$ than $H^\dagger$ to balance hypercharge, which then cannot be made weak isospin invariant. This leaves the only possibility as shown.}, we arrive at the unique operator at odd $d$ dimension,
\begin{eqnarray}
\calO^{d~pr}_{LH}&=&\big[(L^T_p\epsilon H)C (L^T_r\epsilon H)^T\big](H^\dagger H)^{(d-5)/2},
\label{O_LH}
\end{eqnarray}
where $p,~r$ are lepton flavor indices. This is the generalized dim-$d$ Weinberg operator for Majorana neutrino mass whose uniqueness was established previously in Ref.~\cite{Liao:2010ku} using Young tableau.

The lowest-index operators are of interest because their renormalization running under SM interactions is governed at the leading order by their own anomalous dimensions; i.e., they are only renormalized at the next-to-leading order by higher-index operators of the same canonical dimension. This is evident from Eq.~(\ref{chi_gamma}) and the last row in Tables~\ref{tab_dim6} and \ref{tab_dim7}. The uniqueness of the lowest-index operators at each dimension further simplifies the consideration of their renormalization running, which will be taken up in the remaining part of this work. Before that, we make a connection to classification of operators in terms of their holomorphic and anti-holomorphic weights $\omega,~\bar\omega$~\cite{Alonso:2014rga,Cheung:2015aba}. The weights are defined as $\omega(\calO)=n(\calO)-h(\calO),~\bar{\omega}(\calO)=n(\calO)+h(\calO)$
for an operator $\calO$, where $n(\calO)$ is the minimal number of particles for on-shell amplitudes that the operator $\calO$ can generate and $h(\calO)$ the total helicity of the operator. The claim is that our lowest-index operators $\calO_H^d,~\calO_{LH}^d$ are also the ones with the largest weights, i.e., both of their $\omega$ and $\bar\omega$ are the largest among operators of a given canonical dimension. To show this, we introduce some notations. We denote $\Psi$ to be left-handed fermion fields, i.e., $L,~Q,~e^C,~u^C,~d^C$, and $\bar\Psi$ the right-handed ones, and $X^{\mu\nu}_\pm=X^{\mu\nu}\mp(i/2)\epsilon^{\mu\nu\rho\sigma}X_{\rho\sigma}$. The pair of weights has the values $(\omega,\bar\omega)=(1,1),~(1,1),~(3/2,1/2),~(1/2,3/2),~(0,0),~(0,2),~(2,0)$ for the building blocks of operators, $H,~H^\dagger,~\Psi,~\bar\Psi,~D,~X_-,~X_+$, respectively. The weights $(\omega(\calO^d),\bar\omega(\calO^d))$ of an operator $\calO^d$ of dimension $d$ are the sum of the corresponding weights of its components:
\begin{eqnarray}
\omega(\calO^d)&=&n_H+n_{H^\dagger}+\frac{1}{2}(n_{\Bar\Psi}+3n_\Psi)+2n_{X_+}
=d-(n_{\bar\Psi}+n_D+2n_{X_-})\leq d,
\\
\bar\omega(\calO^d)&=&n_H+n_{H^\dagger}+\frac{1}{2}(3n_{\Bar\Psi}+n_\Psi)+2n_{X_-}
=d-(n_{\Psi}+n_D+2n_{X_+})\leq d,
\end{eqnarray}
where $n_B$ denotes the power of the component $B$ appearing in $\calO^d$. The largest $\omega$ and $\bar\omega$ that an operator could have is thus its canonical dimension. For $d$ even, this is easy to realize by sending $n_{X_\pm}=n_D=n_\Psi=n_{\bar\Psi}=0$, i.e., the operator with the highest weights is the lowest-index operator $\calO_H^d$ made up purely of the Higgs field. For $d$ odd, it is known that all operators in SMEFT necessarily involve fermion fields \cite{Degrande:2012wf}, with the minimal choice being $n_\Psi+n_{\bar\Psi}=2$. This can be arranged by choosing $n_\Psi=2,~n_{X_\pm}=n_D=n_{\bar\Psi}=0$ resulting in the operator $\calO_{LH}^d$ of the highest weights $(d,d-2)$, or by choosing instead $n_{\bar\Psi}=2$ as its Hermitian conjugate $\calO_{LH}^{d\dagger}$. The alternative choice $n_\Psi=n_{\bar\Psi}=1$ would require a factor of $D$ due to Lorentz symmetry, which reduces $\omega$ (or $\bar\omega$) by two units compared with $\calO_{LH}^d$ (or $\calO_{LH}^{d\dagger}$). This establishes the claim. As a side remark, the above equations together with Lorentz symmetry also imply that the operators at even (odd) dimension have even (odd) holomorphic and anti-holomorphic weights.

\begin{figure}[!htbp]
\begin{center}
\includegraphics[width=0.6\linewidth]{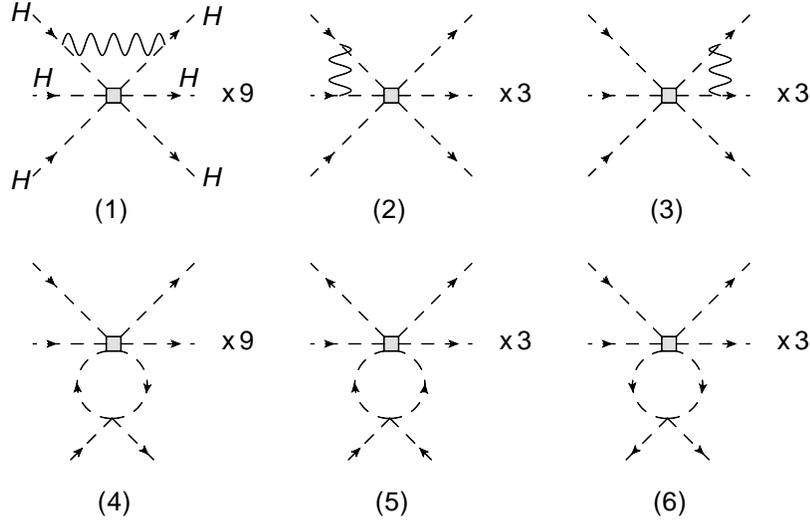}
\end{center}
\caption{One-loop Feynman diagrams for self-renormalization of the operator $\mathcal{O}^6_{H}$ shown as a grey square. The wavy (dashed) line represents gauge (scalar) fields. The arrows indicate the flow of hypercharge.}
\label{fig1}
\end{figure}

\begin{figure}[!htbp]
\begin{center}
\includegraphics[width=0.6\linewidth]{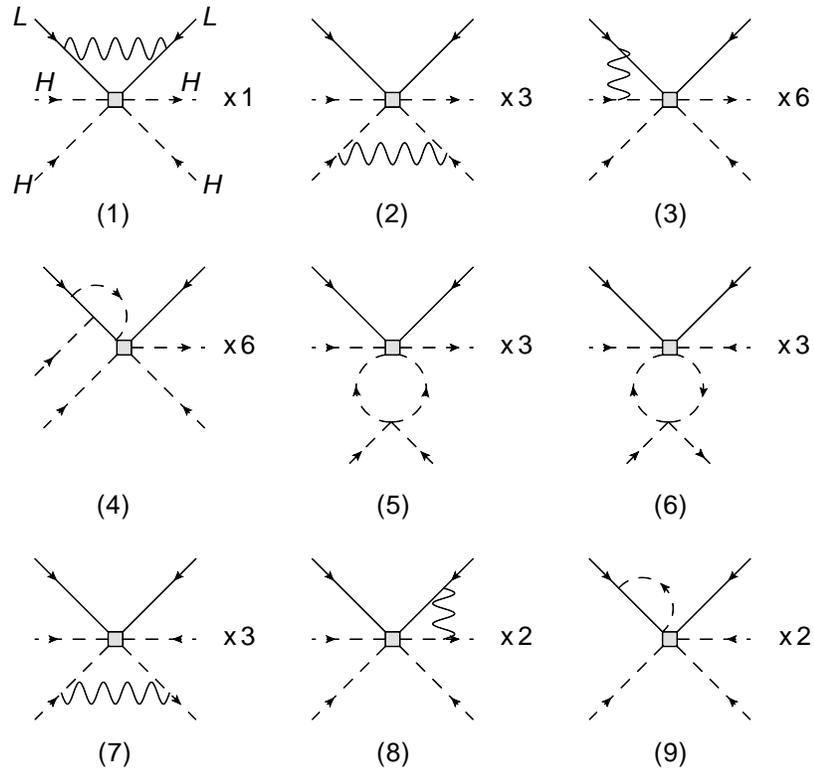}
\end{center}
\caption{
Similar to Fig.~\ref{fig1} but for the operator $\calO_{LH}^{7~pr}$. The arrow on the solid line stands for lepton number flow. For $d\geq 9$ the incoming scalars to the grey square in the loop of graphs (2) and (5) can also be outgoing.
}
\label{fig2}
\end{figure}

Now we compute the anomalous dimensions at leading order for the lowest-index operators $\calO_H^d$ at even dim-$d$ and $\calO_{LH}^{d~pr}$ for odd dim-$d$ in Eqs.~(\ref{O_H},\ref{O_LH}). The Feynman diagrams shown in Figs.~\ref{fig1} and \ref{fig2} are for $\calO_H^6$ and $\calO_{LH}^{7~pr}$ respectively. At higher dimensions one has to be careful with combinatorics due to powers of $H^\dagger H$ involved in the operators. We perform the calculation in dimensional regularization and minimal subtraction scheme and in the general $R_\xi$ gauge. The cancelation of the $\xi$ parameters in the final answer then serves as a useful check. The renormalization group equations for the Wilson coefficients of the above two operators are, at leading order in perturbation theory,
\begin{eqnarray}
\label{rge_even}
16\pi^2\mu\frac{d}{d\mu}C^d_H&=&
\left[3d^2\lambda -\frac{3}{4}dg_1^2 -\frac{9}{4}dg_2^2 +dW_H \right]C^d_H,
\\
\nonumber
16\pi^2\mu\frac{d}{d\mu}C^{d~pr}_{LH}&=&
\left[(3d^2-18d+19)\lambda -\frac{3}{4}(d-5)g_1^2 -\frac{3}{4}(3d-11)g_2^2 +(d-3)W_H\right]C^{d~pr}_{LH}
\\
&&-\frac{3}{2}\left[(Y_eY^\dagger_e)_{vp}C^{d~vr}_{LH}+(Y_eY^\dagger_e)_{vr}C^{d~pv}_{LH}\right],
\label{rge_odd}
\end{eqnarray}
where $W_H={\Tr}[3(Y^\dagger_uY_u)+3(Y^\dagger_dY_d)+(Y^\dagger_eY_e)]$ comes from field strength renormalization of $H$.

We make some final comments on the above result. The terms in the anomalous dimensions due to the Higgs self-coupling $\lambda$ increase quadratically with canonical dimension $d$ due to combinatorics, making renormalization running effects significantly more and more important for higher dimensional operators. The Yukawa terms in Eq.~(\ref{rge_odd}) are independent of $d$ because the lepton field $L$ cannot connect to $(H^\dagger H)^{(d-5)/2}$ to yield a nonvanishing contribution due to weak isospin symmetry. The large numerical factor in the $\lambda$ term for $C_H^6$ was observed previously in \cite{Jenkins:2013zja}, and our leading order results indeed match that work. Including a symmetry factor of $1/2$ in the $\lambda$ term of Eq.~(\ref{rge_even}) that appears in graphs (4)-(5) in Fig.~\ref{fig1} at $d=4$, our result also applies to renormalization of the $\lambda$ coupling and is consistent with \cite{Arason:1991ic} upon noting different conventions for $\lambda$. The renormalization of the Weinberg operator $\calO_{LH}^{5~pr}$ was finally given in Ref.~\cite{Antusch:2001ck} and corresponds to graphs (1)-(5) in Fig.~\ref{fig2}. Our result at $d=5$ is consistent with that work again after taking into account different conventions for $\lambda$. The $\lambda$ term of the $\gamma$ function for $\calO_{LH}^{d~pr}$ increases significantly with $d$ for the first two operators in particular, from $4\lambda$ at $d=5$ to $40\lambda$ at $d=7$.

In summary, we have provided a simple perturbative power counting for renormalization effects of higher dimensional operators due to SM interactions in the framework of SMEFT. In the course of our analysis we introduced an index that parametrizes the perturbative order of operators. We found that the lowest-index operators are unique at each mass dimension, and that their renormalization running under SM interactions is determined at leading perturbative order by their own anomalous dimensions. We computed the anomalous dimensions of those operators for any mass dimension and found that they increase quadratically with their mass dimension. This will be useful in the study of effective scalar potential and generation of tiny Majorana neutrino masses in the framework of SMEFT.


\section*{Acknowledgement}

This work was supported in part by the Grants No. NSFC-11025525, No. NSFC-11575089 and by the CAS Center for Excellence in Particle Physics (CCEPP).

\noindent %

\end{document}